\documentstyle[aps,preprint]{revtex}
\begin{document}
\draft
\title{ Supersymmetry in models with strong on-site Coulomb repulsion -
application to t-J model 
}
\author{T. K. Ng and C. H. Cheng}
\address{Dept. of Physics, HKUST, Kowloon, Hong Kong}
\date{ \today }
\maketitle
\begin{abstract}
   A supersymmetric way of imposing the constraint of no double
occupancy in models with strong on-site Coulomb repulsion is 
presented in this paper. In this formulation the physical
operators in the constrainted Hilbert space are  
invariant under local unitary transformations mixing
boson and fermion representations. As an illustration the
formulation is applied to the $t-J$ model. The model is studied
in the mean-field level in the $J=0$ limit where we show
how both the slave-boson and slave-fermion formulations are
included naturally in the present approach and how further
results beyond both approaches are obtained.
\end{abstract} 


\narrowtext
   The $t-J$ model has became a focus in the study of strongly
correlated metals and High-$T_c$ superconductors since it was proposed
in late eighties\cite{and}. Because of lack of small parameters for
expansion, analytical understandings of the model were largely depending on
mean-field theories which treat the constraint of no double occupancy
only on average. So far, the most successful mean-field approaches 
to the $t-J$ model seems to be based on either the slave-fermion
mean-field theory(SFMFT)\cite{sf} which is successful at very small
doping when antiferromagnetic correlation is important, and the
slave-boson mean-field theory(SBMFT)\cite{sb} which is successful
at larger value of doping when the system becomes superconducting.
The only difference between the two approaches is that two
different representations of spin and electron operators are used
to impose the constraint of no double occupancy.
More recently, the focus in the study of High-$T_c$ superconductors
has turned to the underdoped and spin-glass regimes where it is
believed that the subtle interplay between antiferromagnetism and 
superconductivity determines the properties of this crossover
region. In particular, the
importance of $SU(2)$ symmetry in the underdoped regime of the
$t-J$ model has been pointed out\cite{wl,lw2}. Alternatively, it
was also suggested that an $SO(5)$ symmetry may play an
important role in determining the competition between antiferromagnetism
and superconductivity in the high-$T_c$ cuprates\cite{zhang}.

  To understand the complicated behaviour in this regime of the
$t-J$ model, it seems that a unified approach which
incorporate both the advantages of the slave-fermion mean-field theory 
and the slave-boson mean-field theory is essential. In this
paper, we shall show that it is possible in general to formulate
models with constraint of no double occupancy in a way which
incorporates the advantages of both slave-fermion and slave-boson
representations. In this new formulation the physical operators are
{\em supersymmetric} and are invariant under unitary transformations
mixing fermion and boson representations. The formulation suggests 
that supersymmetry exists naturally in
strongly-correlated systems where on-site Coulomb repulsions
are strong. In the following, we shall use the $t-J$ model as
an example to illustrate our approach. To begin with,
we first consider the Hilbert space of a lattice model with 
constraint of no double occupancy imposed.

   The constraint of no double occupancy implies that there are 
three possible states on any single lattice site in the model. The
site can be either empty (hole state), or can be occupied by
either an up- or down- spin electron. In the slave-boson approach,
the hole state is represented as a boson, whereas spins are
represented as fermions\cite{sb}. It is also equally valid
to represent spins as bosons, and holes as slave 
fermions, as in the slave fermion treatment\cite{sf}. In our
formulation we shall consider an enlarged Hilbert space where
both possibility of representing hole and spins coexist as
different states of the system, i.e. there are now
six possible states per site in the Hilbert space, represented by,
\begin{mathletters}
\label{hilbert}
\begin{equation}
\label{hil1}
|\sigma_i^{(f)}>=c^+_{i\sigma}|0>\;, \;\;\;
|h_i^{(b)}>=b^+_i|0>,
\end{equation}
and
\begin{equation}
\label{hil2}
|\sigma_i^{(b)}>=\bar{Z}_{i\sigma}|0>\;, \;\;\;
|h_i^{(f)}>=f^+_{i}|0>,
\end{equation}
\end{mathletters}
where $\sigma=\uparrow,\downarrow$, $c^+_{i\sigma}$ and 
$\bar{Z}_{i\sigma}$ are fermionic and bosonic spin creation
operators, respectively and $|0>$ is the vacuum state. Similarly,
$b^+_i$ and $f^+_i$ are bosonic and fermionic hole creation
operators, respectively. Notice that we have seperated the
states into "slave-boson" (1a) and "slave-fermion" (1b) groups
in Eq.\ (\ref{hilbert}). For a system of $N$-lattice sites,
both groups of states are allowed at all sites in our formulation and
the total Hilbert space is thus $2^N$ times larger than the Hilbert
space of the original model. The essence of our approach is to 
construct a Hamiltonian which is equivalent to the original model
in all these $2^N$ groups of states, and consequently our system
with enlarged Hilbert space is equivalent to $2^N$ replicas
of the original model. To see how this Hamiltonian can be
constructed for the $t-J$ model we first consider spin operators.

 We consider the spin operator $\vec{s}_i$ at site $i$,
\begin{mathletters}
\label{spino}
\begin{equation}
\label{sp1}
\vec{s}_i=\vec{s}^{(f)}_i+\vec{s}^{(b)}_i, 
\end{equation}
where
\begin{equation}
\label{sp2}
\vec{s}^{(f(b))}_i=\left(c^+(\bar{Z})_{i\uparrow},
c^+(\bar{Z})_{i\downarrow}\right)
\vec{\sigma}\left(\begin{array}{r}
c(Z)_{i\uparrow} \\
c(Z)_{i\downarrow}
\end{array}\right),
\end{equation}
\end{mathletters}
and $\sigma$ is the usual Pauli matrix. Notice that
$\vec{s}^{(f)}$ and $\vec{s}^{(b)}$ are the spin operators
in usual slave-boson and slave-fermion representations, respectively.
It is obvious that $\vec{s}_i$ is itself a spin
operator since it is the sum of two spin operators. The matrix elements
$<\sigma_i^{(\alpha)}|\vec{s}_i|\sigma{'}_i^{(\beta)}>$ where
$\alpha,\beta=f,b$ and $\sigma,\sigma{'}=\uparrow,\downarrow$
can be computed easily where it is easy to see
that $<\sigma_i^{(f)}|\vec{s}_i|\sigma{'}_i^{(f)}>=
<\sigma_i^{(b)}|\vec{s}_i|\sigma{'}_i^{(b)}>$ and gives the usual
spin operator matrix elements between spin-$1/2$ states whereas
all other matrix elements with $\alpha\neq\beta$ are equal to zero.
Thus the states $|\sigma_i^{(f)}>$ and $|\sigma_i^{(b)}>$ together form 
two identical replicas of spin-$1/2$ states on site $i$ with our 
definition of spin operator \ (\ref{spino}). It can then be shown
by direct evaluation that the Hamiltonian
\[
H_J=J\sum_{<i,j>}\vec{s}_i.\vec{s}_j,  \]
represents $2^N$ identical copies of Heisenberg interaction in our
system of enlarged Hilbert space. 

  The electron annihilation and creation operators in our system
can be defined as $\psi_{i\sigma}=
h^+_i\xi_{i\sigma}$, and $\psi^+_{i\sigma}=\xi^+_{i\sigma}h_i$,
respectively, where
\begin{equation}
\label{hxi}
h_i=\left(\begin{array}{r}
f_i \\
b_i
\end{array}\right)\; , \;\;\;
\xi_{i\sigma}=\left(\begin{array}{r}
Z_{i\sigma}  \\
c_{i\sigma}
\end{array}\right),
\end{equation}
are doublets of hole and spin operators carrying fermion and boson
statistics. It is straightforward to show that the electron
operators defined this way satisfies the usual electron
commutation relations in the Hilbert space spanned by
Eq.\ (\ref{hilbert}). The kinetic energy term $H_t$ of the $t-J$
model can be constructed by requiring that the hopping matrix elements
$<\sigma_j^{\alpha'},h_i^{\beta'}|H_t|h_j^{\alpha},\sigma_i^{\beta}>
=-t_{ij}\delta_{\alpha\alpha'}\delta_{\beta\beta'}$, where $\alpha,
\alpha',\beta,\beta'=b,f$ and $|h_j^{\alpha},\sigma^{\beta}>$
represents a state with a hole belonging to group $\alpha$ on site
$j$ and a spin $\sigma$ belonging to group $\beta$ on site $i$.
Notice that the group indices $\alpha, \beta$'s are "conserved"
in constructing $H_t$.
It is straightforward to show that
\begin{equation}
\label{t-term}
H_t=-t\sum_{<i,j>,\sigma}\left(c^+_{j\sigma}b_jb^+_ic_{i\sigma}+
c^+_{j\sigma}b_jf^+_iZ_{i\sigma}
+\bar{Z}_{j\sigma}f_jb^+_ic_{i\sigma}
-\bar{Z}_{j\sigma}f_jf_i^+Z_{i\sigma}+c.c.\right),
\end{equation}
where we have considered a Hamiltonian with nearest neighbor hopping
only. The four different terms in $H_t$ 
give the matrix elements of the hopping term between the four
possible combination of groups of states at sites $<i,j>$. With
this it is easy to verify that our system with Hamiltonian $H=H_J+H_t$ 
is equivalent to $2^N$ copies of the usual $t-J$ model.
Notice that the hopping term cannot be simply represented as 
$H_t=-t\sum(\psi^+_{i\sigma}\psi_{j\sigma}+c.c.)$ 
because of the sign difference in the "slave-fermion"  term
$\bar{Z}_{j\sigma}f_jf_i^+Z_{i\sigma}$. This sign difference
is well known in studies of slave-fermion mean-field 
theory\cite{sf}.

  The invariance of our system under change of group of states
defined in Eq.\ (\ref{hilbert}) at any lattice site can be
expressed in the language of supersymmetry. To see that first we
note that the spin operator \ (\ref{spino}) can be written using
the $\xi_{\sigma}$ fields as
\begin{equation}
\label{spinxi}
s_i^z ={1\over2}(\xi^+_{i\uparrow}\xi_{i\uparrow}-\xi^+_{i\downarrow}
\xi_{i\downarrow})\;, \;\;\;
s^+_i = \xi^+_{i\uparrow}\xi_{i\downarrow}\; , \;\;\;
s^-_i = \xi^+_{i\downarrow}\xi_{i\uparrow},
\end{equation}
where $\xi^+_{i\sigma}\xi_{i\sigma'}=
\bar{Z}_{i\sigma}Z_{i\sigma'}+c^+_{i\sigma}c_{i\sigma}$.
  It is now easy to see that the spin and electron operators
are invariant under the super-unitary transformation
\begin{equation}
\label{un}
\xi_{i\sigma}\rightarrow{U}_i\xi_{i\sigma} \;,  \;\;\;
h_i\rightarrow{U}_ih_i
\end{equation}
where $U_i$'s are local $2\times2$ unitary super-matrices mixing the fermion 
and boson representations of spins and holes. The generators of the
super-unitary transformations are superspin operators
\begin{eqnarray}
\label{gen}
S^z_i & = & {1\over2}\left((\sum_{\sigma}c^+_{i\sigma}c_{i\sigma}+b^+_ib_i)
-(\sum_{\sigma}\bar{Z}_{i\sigma}Z_{i\sigma}+f^+_if_i)\right),
\\ \nonumber
S^+_i & = & \sum_{\sigma}c^+_{i\sigma}Z_{i\sigma}-b^+_if_i,
\\ \nonumber
S^-_i & = & \sum_{\sigma}\bar{Z}_{i\sigma}c_{i\sigma}-f^+_ib_i,
\end{eqnarray}
where we also define the (magnitude$)^2$ of the superspin as
$S^2=S^zS^z+{1\over2}(S^+S^-+S^-S^+)$.
Using the fact that the allowed states in our Hilbert space
can only be singly ocupied by either spin or hole, it is easy
to show that
\begin{equation}
\label{cas}
S_i^2={3\over4}\left(\sum_{\sigma}\xi^+_{i\sigma}\xi_{i\sigma}
+h^+_ih_i\right)={3\over4},
\end{equation}
where we have used the requirement that $\sum_{\sigma}(\bar{Z}_{i\sigma}
Z_{i\sigma}+c^+_{i\sigma}c_{i\sigma})+f^+_if_i+b^+_ib_i=1$ 
in our Hilbert space \ (\ref{hilbert}) in writing down the last equality.
Notice that the constraint of no double occupancy is 
equivalent to the condition that the
{\em magnitudes} of the superspins are fixed ($=1/2$) 
on all lattice sites! In terms of the superspin operators, the
slave-boson and slave-fermion representations are equivalent to 
fixing the direction of superspins to be pointing up and down,
respectively and supersymmetry expresses the fact that the physical
observables in the system are in fact, invariant under local 
(but time-independent) rotations in the superspin space. 
Notice that because of the "minus" sign in
the "slave-fermion" hopping term $\bar{Z}_{j\sigma}f_jf^+_iZ_{i\sigma}$, 
supersymmetry is straightly speaking, broken by the $t-$ term in the
$t-J$ model. However, our formulation of $t-J$ model in the enlarged
Hilbert space \ (\ref{hilbert}) does not require supersymmetry in
the Hamiltonian and is still valid.

 The Lagrangian of the $t-J$ model in the enlarged Hilbert space is
\begin{eqnarray}
\label{lag}
L & = & i\hbar\sum_{i,\sigma}\xi_{i\sigma}^+{\partial\over\partial{t}}
\xi_{i\sigma}+i\hbar\sum_ih^+_i{\partial\over\partial{t}}h_i-H_t-H_J
\\ \nonumber
& & +\sum_i\lambda_i\left(\sum_{\sigma}\xi^+_{i\sigma}\xi_{i\sigma}+
h^+_ih_i-1\right)-\mu\sum_ih^+_ih_i,
\end{eqnarray}
and a Path-integral formulation of the problem can be written down
as usual. In particular, the partition function of the present model
is $2^N$ times the partition function of the original $t-J$ model.
Notice that the constraint of no double
occupancy in our enlarged Hilbert space is imposed by an
Lagrange multiplier term as usual.

   It is obvious that the same approach can be applied to other
models with constraint of no double occupancy, as long as a proper 
Hamiltonian can be constructed in the enlarged Hilbert space. For 
example, we find that the Hamiltonian for the Infinite-$U$ Anderson
model in the supersymmetric representation is,
\begin{equation}
\label{and}
H_{and}=\sum_{\vec{k},\sigma}\epsilon_{\vec{k}}a^+_{\vec{k}\sigma}
a_{\vec{k}\sigma}+\epsilon_o\sum_{i,\sigma}\xi^+_{i\sigma}\xi_{i\sigma}
+V\sum_{i,\sigma}(a^+_{i\sigma}\psi_{i\sigma}+\psi^+_{i\sigma}a_{i\sigma}),
\end{equation}
where $a(a^+)_{\vec{k}\sigma}$ is the annihilation(creation) operator for 
conduction electrons with momentum $\vec{k}$ and spin $\sigma$ and 
$\xi(\xi^+)_{i\sigma}$ are annihilation(creation) operator for localized
spins on site $i$ where constraint of no double occupancy is imposed.
$\epsilon_{\vec{k}}$, $\epsilon_o$ and $V$ have their usual meaning in
Anderson model. The spin and electron operators in the localized orbitals
are represented by $\xi_{i\sigma}$ and $\psi_{i\sigma}$ operators 
which are supersymmetric in
our formulation and the constraint of no double occupancy can be
imposed by Langrange multiplier fields as
in the $t-J$ model. Notice that the Infinite-$U$ Anderson model is,
in fact, completely supersymmetric, in constrast to the $t-J$ model
where supersymmetry is broken by $t-$ term.

  To see the advantage of present formulation we shall consider
in the following the $t-J$ model in the $J=0$ limit, and
shall study the model at two
dimension in the mean-field level. The model is equivalent to the
$U=\infty$ Hubbard model which is a model of interests in
itself\cite{u1,u2}. 

  A mean-field theory in the supersymmetric $t-(J=0)$-model can be obtained
by making the following decouplings,
\begin{eqnarray}
\label{ct}
H & \rightarrow & -t\sum_{<i,j>,\sigma}\left(<c^+_{j\sigma}c_{i\sigma}>
b_jb^+_i+<b_jb^+_i>c^+_{j\sigma}c_{i\sigma}-<c^+_{j\sigma}c_{i\sigma}>
<b_jb^+_i>+c.c.\right.  \\  \nonumber
& & +<c^+_{j\sigma}Z_{i\sigma}>b_jf^+_i+c^+_{j\sigma}Z_{i\sigma}<b_jf^+_i>
-<c^+_{j\sigma}Z_{i\sigma}><b_jf^+_i>+c.c.   \\  \nonumber
& & +f_jb^+_i<\bar{Z}_{j\sigma}c_{i\sigma}>+<f_jb^+_i>\bar{Z}_{j\sigma}
c_{i\sigma}-<f_jb^+_i><\bar{Z}_{j\sigma}c_{i\sigma}>+c.c. \\  \nonumber
& & \left.-<\bar{Z}_{j\sigma}Z_{i\sigma}>f_jf^+_i-\bar{Z}_{j\sigma}Z_{i\sigma}
<f_jf^+_i>+<\bar{Z}_{j\sigma}Z_{i\sigma}><f_jf^+_i>+c.c.\right) \\ \nonumber
& & +\lambda\sum_{i,\sigma}(\bar{Z}_{i\sigma}Z_{i\sigma}+
c^+_{i\sigma}c_{i\sigma})+(\lambda-\mu)\sum_i(f^+_if_i+b^+_ib_i)
\end{eqnarray}
where $<A>$ is the expectation value of operator $A$ evaluated with the 
mean-field Hamiltonian. The mean-field parameters $\lambda$ and $\mu$ 
are chosen so that
$<f^+_if_i+b^+_ib_i>=\delta$ (hole concentration) and $\sum_{\sigma}
<\bar{Z}_{i\sigma}Z_{i\sigma}+c^+_{i\sigma}c_{i\sigma}>=1-\delta$.
We shall be interested at translationally invariant
solutions where the mean-field parameters are independent of $<i,j>$
in the following. 

   Using the fact that $<C>=0$ for Grassman variables $C$'s, we find that
the second and third lines in the mean-field Hamiltonian \ (\ref{ct})
are equal to zero. However, both slave-boson and slave-fermion type
terms are still present in the mean-field Hamiltonian (first and
fourth lines in \ (\ref{ct})). The relative weight of the two kinds of
mean-field terms are determined by the occupation numbers of
spins (and holes) with fermion (boson) and boson (fermion)
statistics. These numbers are in term, determined by minimizing
the free energy of the system with the constraint that the
{\em total} number of spins and holes are fixed to be
$1-\delta$ and $\delta$, respectively.

   Solving the mean-field equations we find that at
low temperatures there are two solutions corresponds to 
local minima in the mean-field free energy for any 
given hole concentration $\delta$. The two solutions have either
$<c^+c>,<b^+b>\neq0$ and $<\bar{Z}Z>,<f^+f>=0$ or the other way around 
and corresponds to the usual slave-boson and slave-fermion solutions.
New solutions where both slave-fermion and slave-boson mean-field
parameters are nonzero are also present. However, these solutions
correspond to saddle points in free energy landscape and are unstable.
Comparing the free energies of the two stable solutions we find that
at zero temperature and at low doping $\delta\leq0.33$ the
slave-fermion-like solution has lower energy
whereas for $\delta\geq0.33$ slave-boson-like 
solution has lower energy. The slave-fermion 
solution where the spinons condensed with $<Z_{i\uparrow}>=<Z_{i
\downarrow}>$ corresponds to {\em ferromagnetic} state with spin pointing
in $x-$ direction whereas the slave-boson solutions
correspond to paramagnetic state. The mean-field theory predicts a
first-order transition at zero temperature from ferromagnetic
to paramagnetic state as concentration of hole increases across
$\sim0.33$. Notice that numerical\cite{u1} and analytical\cite{u2}
works have established that the Nagaoka (ferromagnetic) state is
unstable in the infinite-U Hubbard model for any finite concentration
of doping $\delta$ and our mean-field phase diagram is incorrect.
Nevertheless, our mean-field theory does
produce the qualitative features that the
spins (holes) excitations are bosonic (fermionic) like in the
ferromagnetic state of the model, and are fermionic (bosonic)
like in the paramagnetic state\cite{u3}.
  
  It has to be emphasized that although the mean-field solutions are
either slave-boson or slave-fermion like, spins and holes excitations
with both statistics are present in the supersymmetric
mean-field theory. For example, in the slave-boson-like
solution where $<\bar{Z}Z>,<f^+f>=0$, bosonic spin and fermionic
hole excitations still appear as dispersionless high
energy excitations with energies $\lambda$ and $\lambda-\mu$,
respectively in the mean-field theory. Similar situation occurs
also in the slave-fermion-like solution. The simultaneous
appearance of spin and hole excitations with both statistics
in the supersymmetric mean-field theory implies that at 
finite temperature, the physical properties of the theory
is quite different from usual slave-boson or slave-fermion
mean-field theories. For example, extra incoherent parts in the
one-electron Green's function which does not exist in usual slave-boson
or slave-fermion theories will appear in the supersymmetric theory
at nonzero temperature.

  At high temperature $T>>\delta(1-\delta){t}$ the only stable mean-field
solution which exists is with mean-field parameters
$<\bar{Z}Z>,<f^+f>,<c^+c>,<b^+b>$ all equal to zero, indicating
that the high-temperature phase is completely incoherent.
The transition from coherent low temperature states to incoherent 
high temperature state is a first-order phase transition in mean-field
theory. However, similar results were also obtained in conventional
slave-fermion or slave-boson mean-field theories where it is 
believed that the prediction of first-order phase transition from
low-temperature to high-temperature phases is a defect of
mean-field theory and should be replaced by a smooth
crossover when temperature increases. The mean-field
phase diagram is shown in figure 1, where the three stable
phases are seperated by lines indicating first-order
phase transitions in mean-field theory.

   Summarizing, by extending the size of Hilbert space, we
present in this paper a new way of formulating $t-J$ model where 
supersymmetry is inherent in the physical observables.
The consequence of this hidden supersymmetry is exploited in the $t-J$ 
model where a new, supersymmetric mean-field theory of the model is introduced
and is studied in the $J\rightarrow0$ limit.
The new mean-field theory unifies the slave-boson and slave-fermion
mean-field treatments of $t-J$ model by including them as subsets of possible
solutions of the new mean-field theory. More generally, new phases where
both slave-boson and slave-fermion type mean-field parameters are
nonzero may also exist.

  Perhaps the more important message in this paper is the demonstration
of existence of supersymmetry in general strongly correlated systems where the
low energy physics is described by effective Hamiltonians with constraint
of no double occupancy like the $t-J$ or infinite-$U$ Anderson model. 
The consequence of this supersymmetry is never exploited in studies of
strongly-correlated systems. Note that supersymmetry is broken
by the $t-$ term in the $t-J$ model and it's consequences are
not fully exploited in our paper presenting only the mean-field
treatment of $t-(J=0)$ model. Further studies of the $t-J$ model is
underway and the results will be presented in future papers.

 We acknowledges the support of HKRGC through grant no. HKUST6143/97P.

\begin{figure}
\caption{phase diagram of the $t_(J=0)$ model in supersymmetric mean-field
theory: (HT)-high temperature phase, (SB)-slave-boson-like phase, (SF)-
slave-fermion-like phase}
\end{figure}
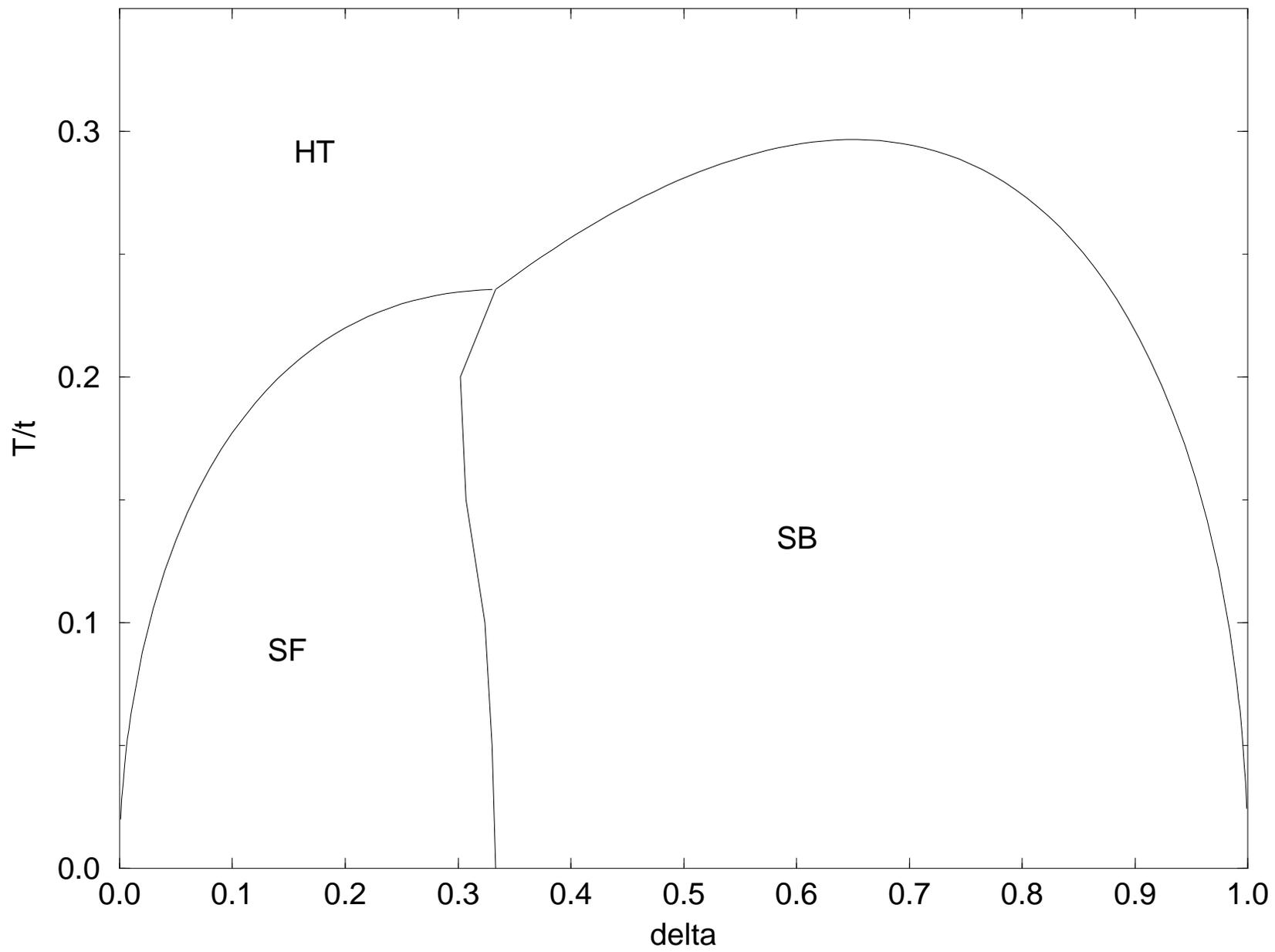
\end{document}